\documentclass[prl,aps,twocolumn,floatfix]{revtex4}
\usepackage{graphicx, amsmath}
\begin{document}

\title{Understanding the $\frac{5}{2}$ Fractional Quantum Hall Effect without the Pfaffian Wave Function}
\author{Csaba T\H oke}
\author{Jainendra K.\ Jain}
\affiliation{Department of Physics, 104 Davey Laboratory, The Pennsylvania State University, Pennsylvania, 16802}

\begin{abstract}
It is demonstrated that an understanding of the $\frac{5}{2}$ fractional quantum 
Hall effect can be achieved within the composite fermion theory without 
appealing to the Pfaffian wave function.  
The residual interaction between composite fermions plays a crucial role 
in establishing incompressibility at this filling factor.  
This approach has the advantage of being amenable to 
systematic perturbative improvements, 
and produces ground as well as excited states.  It, however, does not 
relate to non-Abelian statistics in any obvious manner.
\end{abstract}
\maketitle

The $\frac{5}{2}$ fractional quantum Hall effect (FQHE)  \cite{Willett1,Pan},
has received renewed attention of late because of the intriguing possibility 
of its utilization in quantum computation.  The currently most promising 
picture for this FQHE is in terms of ``the Pfaffian state," 
proposed by Moore and Read \cite{Moore1} and Greiter, Wen, and Wilczek \cite{GWW,Greiter2}.
The Pfaffian wave function describes a real-space p-wave-paired BCS wave function 
for a fixed the number of composite fermions (CFs).
The best evidence in favor of the Pfaffian state comes from numerical studies, which 
have shown \cite{Morf1,Rezayi,Scarola02} that for small systems in the 
second Landau level (LL), the Pfaffian wave function has a reasonably good  
overlap with the exact Coulomb ground state. 
The Pfaffian also 
is the exact ground state for a model three-body short-range  
interaction, ${\cal V}_3$ (Eq.~\ref{modelint}).  Exact solutions for this model 
are also available for quasiholes, which have been shown, theoretically, to constitute  
a realization of ``non-Abelian anyons"  
\cite{Moore1,Nayak1,Read1,TSimon}.

This picture, however, is not entirely satisfactory.
It is not known how the Pfaffian wave function, which 
does not contain any variational parameters, 
can be improved for the two body Coulomb interaction.  
The pairing of composite fermions is viewed as arising from an 
instability of the CF Fermi sea \cite{GWW,Greiter2,Scarola1}, but the CF Fermi sea 
is not a limiting case of the Pfaffian wave function.  
No satisfactory quantitative understanding currently exists for the 
excitations of the $\frac{5}{2}$ state; we see evidence below that the 
three-body interaction ${\cal V}_3$ does not capture the qualitative physics of 
the actual excitations of the Coulomb $\frac{5}{2}$ state.  Finally, the actual 
meaning of pairing of composite fermions remains unclear; the 
$\frac{5}{2}$ state has no off-diagonal long-range order, and, in an 
operational sense, it does not appear different from other FQHE states.

These considerations have motivated us to seek another approach for 
describing the physics of the $\frac{5}{2}$ FQHE, on which we elaborate in this Letter.
We still proceed within the CF framework, but without assuming any pairing 
at the outset.  No FQHE occurs at $\nu=\frac{5}{2}$ in a model that 
neglects interactions between composite fermions, which predicts many 
degenerate ground states at this filling factor.  [This is to be contrasted with 
the FQHE at $\nu=n/(2pn\pm 1)$.]
We show below that the residual interaction 
between composite fermions opens a gap to produce an incompressible state.
Furthermore, the results can be improved systematically within a perturbative 
scheme.  This approach produces the ground state as well as low-lying excitations.

Below, the lowest LL is assumed to be full and 
inert; only the half-filled second LL is considered explicitly, and  
full spin polarization of electrons is assumed.  Lengths are measured in 
the units of the magnetic length, $l_B=\sqrt{\hbar c/eB}$, and 
energies in units of $e^2/\epsilon l_B$ ($\epsilon$ is the dielectric constant 
of the host semiconductor). The spherical geometry \cite{Haldane} will be 
employed,  which considers $N$ electrons moving on  
the surface of a sphere with a magnetic monopole of strength $Q$ 
at the center, producing a magnetic flux of strength $2Q\phi_0$, 
where $\phi_0=hc/e$ is the magnetic flux quantum.  The orbital angular 
momentum quantum number is denoted by $L$.

The Pfaffian state assumes the form
\begin{equation}
\label{pfaffsphere}
\Psi^{\text{Pf}}=\text{Pf}\left(\frac{1}{u_iv_j-v_iu_j}\right)\Phi_1^2,\quad\Phi_1=\prod_{i<j}(u_iv_j-v_iu_j)\;,
\end{equation}
where $u_i=\cos\frac{\theta_i}{2}e^{-i\phi_i/2}$ and 
$v_i=\sin\frac{\theta_i}{2}e^{i\phi_i/2}$.  It is the exact ground state of the 
short-range three-body interaction \cite{GWW,Read1,Rezayi} 
\begin{equation}
{\cal V}_3=V\sum_{i<j<k} P_{ijk}(L_{\rm max})
\label{modelint}
\end{equation}
where $P_{ijk}(L_{\rm max})$ is the projection operator onto an electron triplet with
orbital angular momentum $L_{\rm max}=3Q-3$. 
${\cal V}_3$ penalizes configurations 
with electron triplets in their closest configuration.

The CF theory \cite{Jain1} describes the two-dimensional electron 
system in terms of 
composite fermions,  which are electrons bound to an even number
(taken to be two in this paper) of quantized vortices.  The lowest LL splits 
into ``$\lambda$-levels" of composite fermions, which are analogous to 
Landau levels of electrons in a reduced field $B^\ast=B-2\rho\phi_0$.
Microscopically, the CF formation is defined by the expression 
\begin{equation}
\label{mapping}
\Psi_Q = P_{\rm LLL}\Phi_1^{2}\Phi_{Q^\ast},
\end{equation}
where $\Phi$ is a wave function for $N$ electrons at monopole strength $Q^\ast$,
the Jastrow factor $\Phi_1^{2}$ attaches two vortices 
to them, and $P_{\rm LLL}$ projects the wave function into the lowest 
LL \cite{JainKamilla}.
The monopole strengths are related by 
$Q = Q^\ast + N-1$.

A technical obstacle toward a quantitative study of the state at  
$\nu=\frac{5}{2}$, defined here through the relation $2Q=2N-3$,
is that composite fermions experience a negative 
magnetic field here, given by $Q^\ast=-\frac{1}{2}$.  While the CF theory 
is known to be valid for 
negative $B^\ast$ \cite{Wu93,Moller}, the convenient projection method developed in 
Ref. \cite{JainKamilla} does not apply to such situations for technical reasons.
(The recent work of M\"oller and Simon \cite{Moller} can be 
useful in this respect, but we have not explored that.)
We avoid negative values of $Q^\ast$ by exploiting particle-hole symmetry to study 
$N_h=(2Q+1)-N=N-2$ {\em holes} at $2Q=2N-3$.  Composite fermions {\em 
made from holes} experience a positive monopole strength 
\begin{equation}
Q^\ast=Q-(N_h-1)=3/2\;.
\end{equation}
The hole version of $\Psi^{\text{Pf}}$ is found 
conveniently from its second-quantized form.

\begin{table}[htb]
\begin{center}
\begin{tabular}{r|r|r|r|r|r|r|r}
\hline\hline
$N_h$ & $D_{ex}$ & $D^{(0)}$ & $D^{(0)}_{L=0}$ & $D^{(1)}$ & $D^{(1)}_{L=0}$ & $D^{(2)}$ & $D^{(2)}_{L=0}$ \\ 
\hline
6  &      151      &  3 & 1 & 14 & 2 &  42   &  3  \\
8  &     1514      &  3 & 1 & 20 & 1 &  72   &  4  \\
12 &   194668      &  4 & 1 & 37 & 2 & 205   &  8  \\
14 &  2374753      &  8 & 1 & 63 & 3 & 644$^\ast$  & 18$^\ast$ \\
16 &$3\times 10^7$ &  4 & 1 & 52 & 2 & 495$^\ast$  & 14$^\ast$ \\
20 &$5\times 10^9$ &  5 & 1 & 77 & 2 & 965$^\ast$  & 18$^\ast$ \\
\hline\hline
\end{tabular}
\end{center}
\caption{\label{dims} Dimensions of various bases for $N_h$ particles 
at $2Q=2N_h+1$. $D_{ex}$ is the size of the 
Hilbert space in $L_z=0$ sector,  and $D^{(n)}$ is the dimension of the CF
basis incorporating $n^{\rm th}$ order $\lambda$-level mixing.
$D^{(n)}_{L=0}$ is the number of CF states in the $L=0$ sector. 
Asterisks mark the cases where we could not determine the 
number of linearly independent basis states.
}
\end{table}

The single particle states at $Q^\ast$ are monopole 
harmonics \cite{Wu} $Y_{Q^\ast lm}$, where $l$ is the angular momentum and 
$m$ its $z$-component.  The LL index is given by $n=l-Q^*$ ($n\ge 0$).
Independent many-fermion basis states $\Phi$ are Slater determinants 
of $Y_{Q^\ast lm}$'s at $Q^*$, specified by a set $\{l_i,m_i\}$.
In the $n^{\rm th}$ order of ``CF diagonalization"\cite{Mandal}, we collect all 
basis states with at most $n$ units of CF kinetic energy above the minimum:
\[
\left\{
\{\Phi^{(0)}_{\alpha}\},\{\Phi_{\beta}^{(1)}\},\{\Phi^{(2)}_\gamma\},\ldots,\{\Phi^{(n)}_\zeta\}
\right\}\;.
\]
A correlated CF basis at $Q$, of dimension $D^{(n)}$, is obtained through 
Eq.~(\ref{mapping}), 
\[
\left\{
\{\Psi^{(0)}_{\alpha}\},\{\Psi_{\beta}^{(1)}\},\{\Psi^{(2)}_\gamma\},\ldots,\{\Psi^{(n)}_\zeta\}
\right\},
\]
with $n_i=l_i-Q^\ast$ now interpreted as the $\lambda$-level index, and 
$\sum_in_i$ as the total ``CF kinetic energy.''  
We diagonalize the Coulomb interaction $V$ in this basis.
That requires a Monte Carlo evaluation of the direct product and 
interaction matrices ($\langle\Psi^{(n)}_\alpha|\Psi^{(m)}_\beta\rangle$ and
$\langle\Psi^{(n)}_\alpha|V|\Psi^{(m)}_\beta\rangle$, respectively), orthogonalization by the standard Gram-Schmidt procedure, and numerical diagonalization\cite{Mandal}.
The ground state from the $n^{\rm th}$ order CF diagonalization
will be denoted by $\Psi_0^{(n)}$.  The dimensions of various bases are 
given in Table \ref{dims}.

Monte Carlo CF diagonalization requires a real space interaction.
The Coulomb interaction of the second LL
is simulated by an effective interaction in lowest LL of the form
\begin{equation}
\label{form}
V^{\text{eff}}(r)=\frac{1}{r}+\sum_{i=0}^M c_i r^i,
\end{equation}
where the coefficients $c_i$ are fixed so that the 
lowest LL pseudopotentials \cite{Haldane} of $V^{\text{eff}}(r)$
reproduce \emph{all} of the second LL Coulomb 
pseudopotentials $V^{(1)}_m$ for odd integral values of $m$.  The 
Coulomb pseudopotentials in the $n^{\rm th}$ LL are given by 
\begin{multline}
V^{(n)}_m=\frac{1}{R}\sum_{m_1,m_2=-l}^{l}\sum_{j=|m_1-m_2|}^{2l}(-1)^{j+m_2-m_1}\times\\
\langle 2l-m,0\mid l,m_1;l,-m_1\rangle
\langle 2l-m,0\mid l,m_2;l,-m_2\rangle\times\\
\left|\langle l,m_1;j,m_2-m_1\mid l,m_2\rangle\langle l,Q;j,0\mid l,Q\rangle\right |^2,
\label{pseudo}
\end{multline}
where $m$ is the relative angular momentum of two particles, 
$l=Q+n$, and $R=\sqrt{Q}\,l_B$ is the radius.
(Eq.~\ref{pseudo} reduces to the expression in Fano \emph{et al.} \cite{Fano} for $n=0$.)
As $Q$ depends on $N$, a distinct set of coefficients has to be calculated for each $N$.
To fit Eq. (\ref{form}) we use that the pseudopotentials of a monomial $r^n$ in the lowest LL are ($J=2l-m$):
\begin{eqnarray}
\label{CCC}
V_m(r^n)&=&\frac{2^{n+4}\pi^2}{(2Q+n/2+1)!(2J+1)!}\times\\
&&\sum_{k=0}^{J}\frac{(J!)^2(J+k)!(2Q+n/2-k)!}{k!(J-k)!}\frac{1}{R}.\notag
\end{eqnarray}

\begin{figure*}[htbp]
\begin{center}
\includegraphics[scale=0.6]{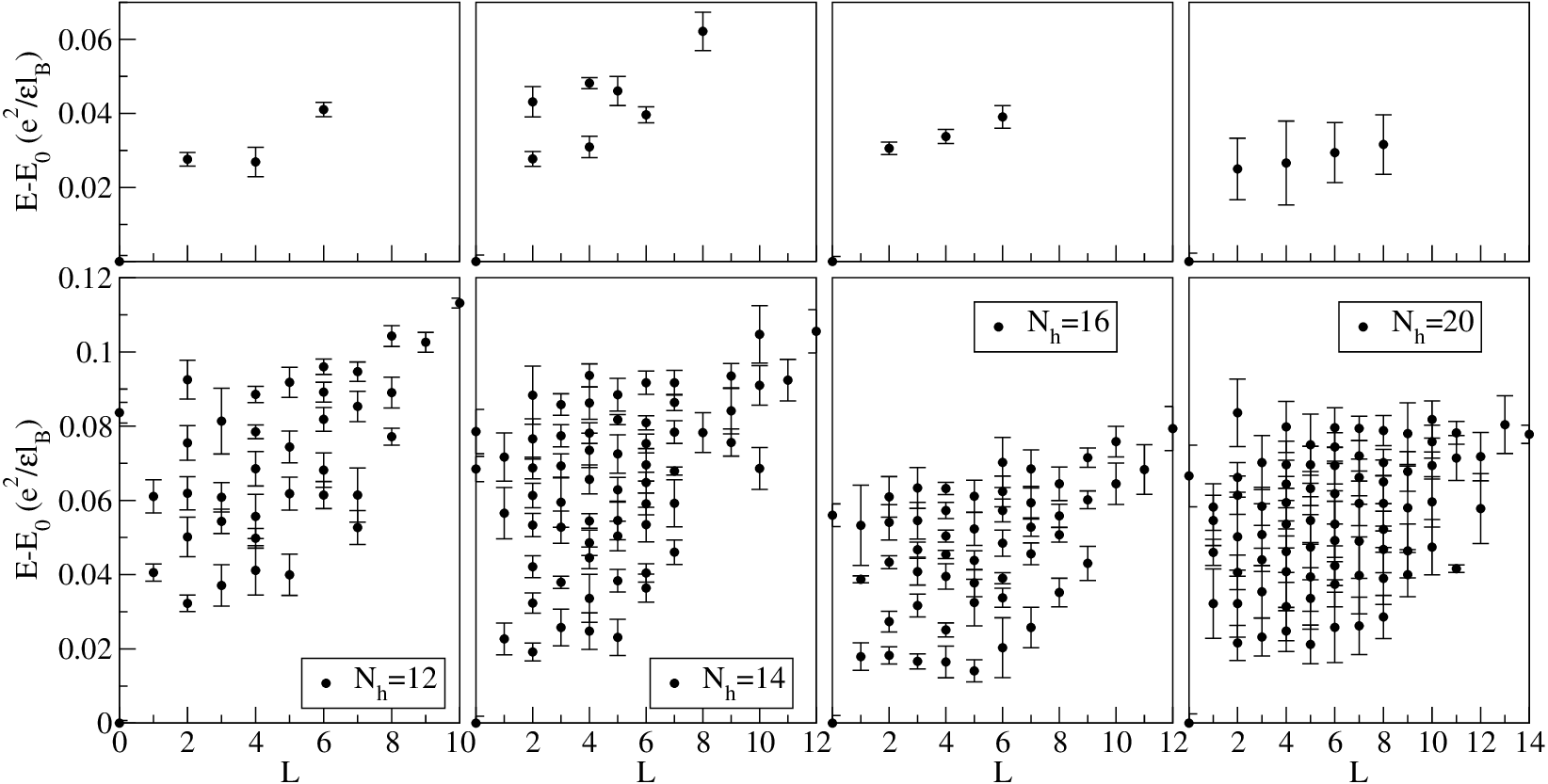}
\end{center}
\caption{\label{specrest} Zeroth-order (top) and first-order (bottom) CF diagonalization excitation spectra for $N_h=12,14,16,20$.
}
\end{figure*}

Figure \ref{specrest} shows the excitation spectra for $N_h=12$, 14, 16 and 20 
obtained by CF diagonalization at the zeroth and the first orders.  
($N_h=18$ is not considered as it aliases with $\nu=3/7$ of holes.)
The residual interaction between 
composite fermions lifts the degeneracy between various states to produce 
an incompressible state already at the lowest (zeroth) order, which neglects 
$\lambda$-level mixing.  Although the energy gaps change by up to 
50\% in going from the the zeroth to the first order, the incompressibility is  
preserved, indicating that 
while $\lambda$-level mixing renormalizes composite fermions, it 
does not cause any phase transition.  The overestimation of gaps at the zeroth 
order may be attributed to the very small dimensions of the CF basis. All
CF basis states are perturbations of the noninteracting CF Fermi sea, 
making it explicit that a rearrangement of composite fermions near the 
CF Fermi level is responsible for the $\frac{5}{2}$ FQHE.

When plotted as a function of $kl_B=L/\sqrt{Q}$, the lowest energy excitations 
for $14\le N_h\le 20$ (from the first order spectra) 
fall on a more or less continuous curve, which 
indicates that the thermodynamic behavior has been approached for $N_h\ge 14$.
Finite-size effects are non-negligible for $N_h<14$.  For $N_h>20$ the first order 
calculation is not sufficient, and the second order CF diagonalization computationally too time consuming.
Although there is some ambiguity as to which excitation  
is to be identified with the transport gap (corresponding to a far separated 
quasiparticle-quasihole pair),
the existence of an almost flat region allows us to estimate a gap of $\sim 0.02$.
This value is consistent with the earlier results from exact 
diagonalization \cite{Morf1,Morf2}.

\begin{figure}[htbp]
\begin{center}
\includegraphics[width=3in]{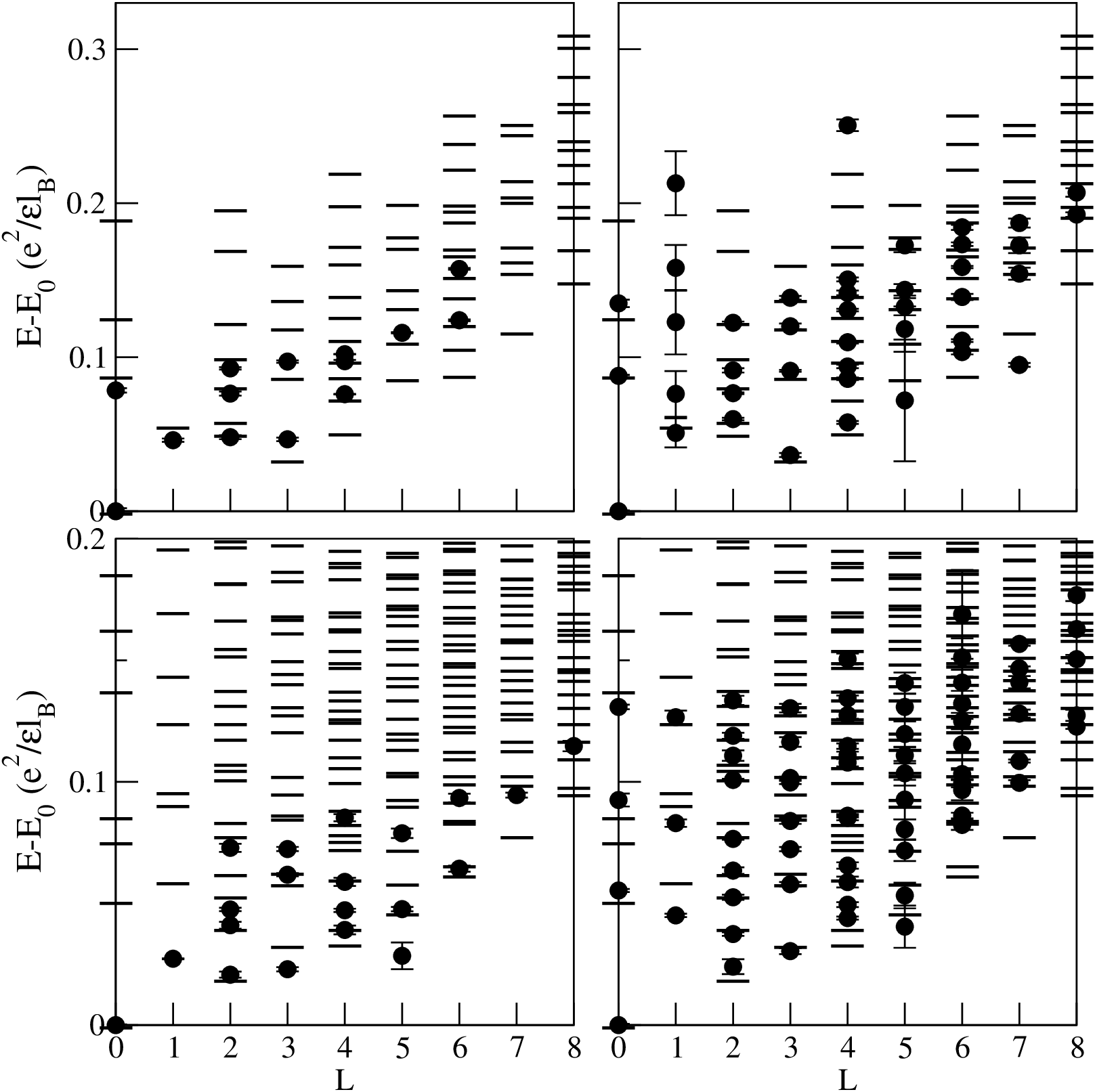}
\end{center}
\caption{\label{cfd6} First-order (left) and second-order (right) CF diagonalization excitation spectra for $N_h=6$ (top) and $N_h=8$ (bottom) holes. The dashes 
show the exact spectrum, and the dots the CF spectrum.  The exact and the CF 
ground state energies for $N_h=6$ are: $E_{ex}/N=-0.415217$, $E^{(1)}/N=-0.413609$, $E^{(2)}/N=-0.415233$; 
those for $N_h=8$ are: $E_{ex}/N=-0.401443$, $E^{(1)}/N=-0.395293$, $E^{(2)}/N=-0.399043$.
}
\end{figure}

Fig.~\ref{cfd6} shows a comparison of the CF spectra, at first and second 
order CF diagonalization, with the exact spectra for $N_h=6$ and 8.
The CF theory does not provide as accurate an account of the energies as 
it does for the lowest LL FQHE states.  However, it works reasonably well for 
energy differences.  The CF spectrum produces, at the first order, the 
energy gap to better than 25\% accuracy.  These comparisons 
thus provide credence to the semi-quantitative validity of our approach.

\begin{figure}[htbp]
\begin{center}
\includegraphics[scale=0.3]{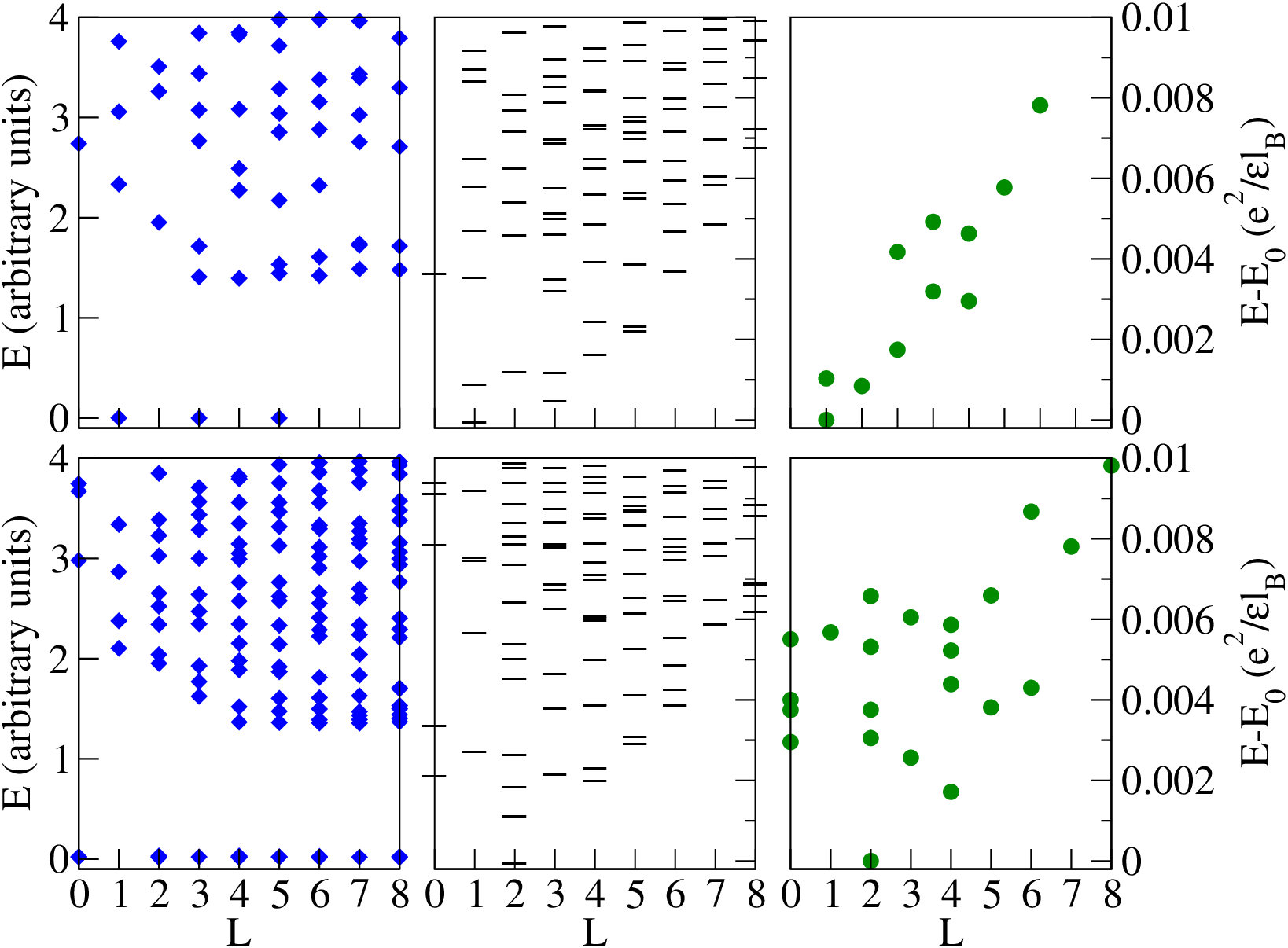}
\end{center}
\caption{\label{comparison}
Spectra at $\nu=5/2$ for the model interaction ${\cal V}_3$ (left column), 
the Coulomb interaction (central column), and the first-order CF
diagonalization (right column) for $N=10$ particles at $2l=18$ (top row) and 
$2l=19$ (bottom row).  For the ${\cal V}_3$ interaction,
two (four) quasiholes are expected for $2l=18$ ($2l=19$).
The ground state energies are $E_0/N=-0.415008$ ($-0.40699$) for exact and $-0.40986$ ($-0.401845$) for composite fermions,
with $2l=18$ ($2l=19$).
The energies in the middle column correspond to the scale shown on right.
The spectra on the left were also given in Ref.~\cite{Read1}.}
\end{figure}

Of interest also is the nature of multi-quasihole states a few flux quanta away 
from $\nu=\frac{5}{2}$.   Fig.~\ref{comparison} shows spectra, for the 
${\cal V}_3$ interaction, for $N=10$ electrons at $2l=18$ and $2l=19$,
which correspond to two and four ``quasiholes" of the Pfaffian state.
(We have switched back to electrons now, as these states occur at positive $B^*$.)
This model predicts zero energy states at $L=1,3,5$ and
 $L=0^2, 1^0, 2^4, 3^1, 4^4, 5^2, 6^3, 7^1, 8^2, 9^0, 10^1$, respectively
(the superscript denotes the degeneracy).  No corresponding quasi-degenerate 
band of states can be identified in the exact spectrum (middle columns).
Fig.~\ref{comparison} also shows spectra from first-order CF diagonalization. 
It produces a ground state at the correct quantum number 
but is not very successful for higher energy states.
The CF spectrum can be improved systematically by incorporating
higher order $\lambda$-level mixing.

The lack of a qualitative correspondence between the low energy 
spectra of ${\cal V}_3$ and the Coulomb interactions in Fig.~\ref{comparison}
raises questions 
regarding the validity of the ${\cal V}_3$ model, and hence of the model of quasiholes 
based on the Pfaffian wave function \cite{Nayak1,Read1}, for the {\em real} quasiholes  
of the $\frac{5}{2}$ state.   This has relevance to the issue of statistics.  
Non-Abelian statistics for the quasiholes of the ${\cal V}_3$ model is a consequence of  
the existence of several degenerate states for a given configuration of spatially localized 
quasiholes, which, in turn, is closely related to the degeneracy of the angular momentum 
eigenstates in Fig.~\ref{comparison}.  The spectra in Fig.~\ref{comparison} 
demonstrate a lack of adiabatic continuity, for the systems studied,
between the many quasihole states of the ${\cal V}_3$ and the Coulomb 
models.  (For the many quasiparticle or many quasihole states of the 
ordinary FQHE states in the lowest 
Landau level, the qualitative structure of a low-energy band predicted by the 
analogy to non-interacting {\em fermions} at $Q^*$ is confirmed in similar exact 
spectra.)

In summary, we have demonstrated that the residual interaction between composite fermions 
causes incompressibility at $\nu=\frac{5}{2}$, and that the lowest 
order treatment of $\lambda$-level mixing gives a reasonable estimate of the 
activation energy.  This model can be applied to neutral excitations at
$\nu=\frac{5}{2}$ as well as charged excitations slightly away from $\nu=\frac{5}{2}$.
The residual interaction may possibly induce pairing between 
composite fermions, but it is not known how to establish that, in a conclusive
manner, within our approach.

The authors thank Chia-Chen Chang for the derivation of Eq.~\ref{CCC}.
We are grateful to him and N.\ Regnault for useful commenta and discussions.
We thank the High Performance Computing (HPC) group at Penn State University ASET 
(Academic Services and Emerging Technologies)
for assistance and computing time on the Lion-XL and Lion-XO clusters.
Partial  support of this research by the National Science Foundation under 
grant No.\ DMR-0240458 is gratefully acknowledged.

\newcommand{\PRL}{Phys.\ Rev.\ Lett.}
\newcommand{\PRB}{Phys.\ Rev.\ B}
\newcommand{\NPB}{Nucl.\ Phys.\ B}

\end{document}